\documentclass[%
preprint,
showpacs,preprintnumbers,
amsmath,amssymb,
aps,
pra,
]{revtex4-1}

\usepackage{graphicx}
\usepackage{dcolumn}
\usepackage{bm}

\usepackage{comment}
\usepackage{color}
\begin{document}


\title{Size-Dependence of the photothermal response of a single metal nanosphere}

\author{Ieng-Wai Un}
\email{iengwai@post.bgu.ac.il}
\affiliation{School of Electrical and Computer Engineering, Ben-Gurion University of the Negev, Israel.}
\affiliation{Joan and Irwin Jacobs TIX Institute, National Tsing Hua University, Taiwan.}
\author{Yonatan Sivan}%
\affiliation{Unit of Electro-optics Engineering, Ben-Gurion University of the Negev, Israel.}



\date{\today}
\begin{abstract}
We study the thermal response of a single spherical metal nanoparticle to continuous wave illumination as a function of its size. We show that the particle temperature increases non-monotonically as the particle size increases, indicating that the photo-thermal response can be optimized by tuning the particle size and illumination wavelength. We also compare the size-effect on the photo-thermal effects of gold and silver nanoparticles and find somewhat surprisingly that Ag NPs are more efficient heat generators only for sufficiently small sizes. 
These results have importance primarily for application such as plasmon-assisted photo-catalysis, photothermal cancer therapy etc., and provide a first step toward the study of the size-dependence of the thermo-optic nonlinearity of metal nanospheres.


\end{abstract}


\maketitle

\section{Introduction}\label{sec:introd}
Metal nanoparticles (NPs) have been studied extensively for the last few decades because of their ability to confine and enhance the electromagnetic field to a sub-wavelength scale. They have found a wide variety of applications~\cite{Brongersma_NatMat_2010,Giannini_chem_rev_2011} such as optical devices in nanoscale~\cite{nano_opt_devices_Adv_Mat_2001}, optoelectronics~\cite{pht_phys_chem_metal_NP_JPCB_2002}, biological sensing~\cite{metal_NP_bio_app_acc_chem_res_2008,Au_NP_bio_sen_chem_rev_2012}, and biomedicine~\cite{Dykman_Acta_Nat_2011,Au_NP_bio_med_2012}. In addition to the good field enhancement and scattering properties, plasmonic nanoparticles have been shown to be ideal heating nanosources when subjected to illumination at their plasmonic resonance wavelength~\cite{Govorov-NP-heat-generation-NanoToday-2007,Baffou-ACSNano-plasmon-heating-shape-2010,Baffou-Quidant-Thermo-plasmonics-review-2013,Baffou-ACSNano-NP_array_heating-2013}. This active research field based on the Joule effect of metal NPs under illumination is usually referred to as thermo-plasmonics~\cite{Govorov-NP-heat-generation-NanoToday-2007,Baffou-Quidant-Thermo-plasmonics-review-2013} and has led to a wide range of emerging applications, especially in biology and energy harvesting, e.g., photo-thermal imaging~\cite{Shaked_LSA_2015,phtthm_imag_science_2002,phtthm_meth_analy_chem_2008}, photothermal therapy~\cite{Shaked_LSA_2015,Cortie_AuNP_hyperthermia_2018,Au_NP_PTT_review_2017,Au_NP_PTT_review_2019}, thermo-photovoltaics~\cite{Zubin_high_temp_ENZ_thmphtvolt_2013,Shalaev_Boltasseva_thm_emi_2015}, plasmonic-heating-induced nanofabrication~\cite{Hashimoto_plasmon_heat_fab_glass_2016,Hashimoto_plasmon_heat_fab_2016}, fluid and molecule dynamics and control in biological applications~\cite{Kall_phtthm_au_nanorod_Brown_ACSNano_2017,Kall_phtthm_DNA_release_ACSpht_2018,Kall_pht_thm_nanoantenna_acspht_2018}, water boiling~\cite{halas_solar_vapor_2013} and plasmon-assisted photocatalysis~\cite{Sivan-Un-Dubi-Faraday,Sivan_Baraban_Un_Dubi_science_comment_2019,Sivan_Un_Dubi_pht_thm_cata_2019,thm_hot_e_faraday_discuss_2019,dyn_hot_e_faraday_discuss_2019,Liu-Everitt-Nano-Letters-2019}.

Under CW illumination, the temperature rise of the NP is determined by the balance between the incoming illumination energy flow and an equal flow of heat to the environment. For sufficiently small particles, it is possible to calculate the absorption by employing the so-called quasi-static approximation. Under this approximation and for fixed illumination intensity, the particle temperature increases monotonically with the particle size. However, as we shall see, the particle temperature varies non-monotonically with the particle size in a manner that was not studied thoroughly so far. Therefore, the present study highlights the importance of the size-dependent particle temperature to identifying the optimal conditions for heat generation. In order to avoid additional complexity due to nonlinear effect~\cite{Sivan-Chu-Nanophotonics-2017}, i.e., the temperature-dependence of the optical and thermal properties of the materials, we limit the illumination intensity such that the particle temperature rise is lower than 100 K such that the material parameters can be assumed to be temperature-independent, as suggested in~\cite{Gurwich-Sivan-PRE-2017}. We focus on modeling the heat generation from a single metal NP. We trace the nature of the non-monotonic behavior to the contributions of the electric dipole and quadrupole modes to the absorption cross-section of the NP. Finally, we compare the size-dependence of photo-thermal response of Au and Ag NPs and identify the particle size and illumination wavelength that yield optimal heat generation. We find, somewhat surprisingly, that while Ag NPs are more efficient heat generators for sufficiently small sizes, Au NPs become more efficient for larger NPs. 

This work is critical for the optimization of the photo-thermal effect in all the above mentioned applications. Most importantly, since thermal effects were shown to be responsible for observations of faster chemical reactions in some of the most famous papers on the topic (see discussion in~\cite{Dubi-Sivan,Sivan-Un-Dubi-Faraday,Sivan_Un_Dubi_pht_thm_cata_2019,Sivan_Baraban_Un_Dubi_science_comment_2019}), our work can be used to interpret correctly the differences in chemical reaction enhancements originating from the use of metal particles of different sizes.

\section{Configuration and Methodology}\label{sec:config}
We consider a {\em single} spherical metallic NP of radius $a$ in a loss-less dielectric host $\epsilon_h$ illuminated by a high intensity CW laser. The absorption of incident photons causes the NP to heat up, an effect which is balanced by heat transfer to the environment such that the temperature reaches a steady state. In this case, the heat equation reduces to the Poisson equation,
\begin{align}\label{eq:poissoneq}
\nabla \cdot \left(\kappa({\bf r}) \nabla T({\bf r}) \right) = - p_{\textrm{abs}}(\omega,{\bf r}), 
\end{align}
where $\kappa$ is the thermal conductivity and $p_{\textrm{abs}}(\omega,{\bf r})$ is the absorbed power density. Here, we only consider one-photon absorption and neglect potential multi-photon absorption so that the time averaged absorbed power density is given by $p_{\textrm{abs}}(\omega,{\bf r}) = \dfrac{\omega}{2} \varepsilon_0 \varepsilon^{\prime\prime}(\omega) |{\bf E}(\omega,{\bf r})|^2$, where ${\bf E}(\omega,{\bf r})$ is the total (local) electric field. The total (local) electric field and, thus, the absorbed power density can be in general obtained by the Mie solution~\cite{bohren_2008_abs_sca_book} to the Maxwell's equations. For very small spherical NPs, one can solve the Maxwell's equations numerically, or based on the quasi-static (QS) approximation~\cite{bohren_2008_abs_sca_book}. 

Since the heat source $p_{\textrm{abs}}(\omega,{\bf r})$ in Eq.~(\ref{eq:poissoneq}) is, in general, spatially-dependent, it has no simple closed-form solution. Thus, we used COMSOL Multiphysics to solve the Maxwell's equations to obtain the electromagnetic fields and $p_{\textrm{abs}}(\omega,{\bf r})$ and then solved Eq.~(\ref{eq:poissoneq}) to obtain the temperature distribution. For solving the Maxwell's equations, a plane-wave illumination was set by using a port boundary condition; perfectly matched layers and scattering boundary conditions were used in the surrounding boundaries in order to eliminate unphysical reflections. For the temperature calculations, we applied transformation optics to the heat equation~\cite{Guenneau-thermo-tranf-OptExpr-2012,Vemuri-Bandaru-heat-meta-APL-2014} and carefully assigned the perfectly matched layers with an anisotropic thermal conductivity to reduce the volume of the simulation domain. This approach is potentially more accessible compared with the volume integral simulation techniques described in~\cite{Sheng_thmplasmon_sim_VIE_MoM_2017}.

\begin{figure}[h]
\centering
\includegraphics[width=1\textwidth]{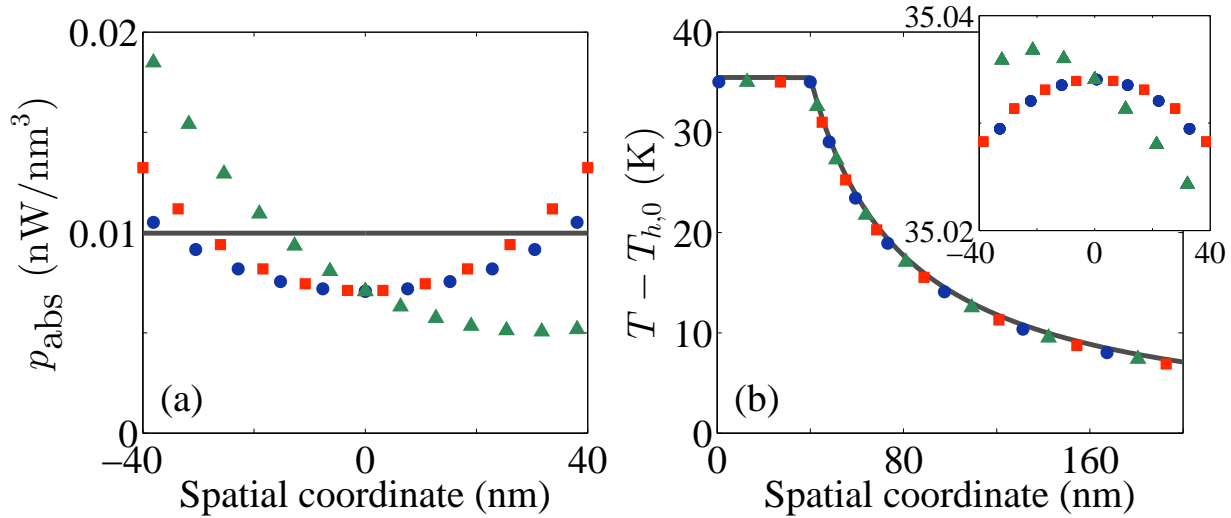}
\caption{(Color online) (a) The absorbed power density and (b) the temperature distribution of a single Au NP (40 nm in radius) in oil under an illumination (y-polarized, incidence along z-axis) with intensity 40 kW/cm$^2$ at wavelength of 532 nm. The black solid line represents the averaged absorbed power density $\bar{p}_{abs}$ in (a) and the approximate analytic Mie-heat solution~(\ref{eq:solveheatgauss}) in (b). The blue-circle, red-square and green-triangle dots represent the COMSOL simulation results along x-, y- and z-axis, respectively. The inset in (b) shows the zoomed-in region of the NP.} \label{fig:temp_mie_vs_comsol_40_532}
\end{figure}

Fig.~\ref{fig:temp_mie_vs_comsol_40_532} shows the COMSOL simulation results of the absorbed power density and the temperature distribution of a single Au NP of 40 nm in radius subjected to CW illumination at wavelength 532 nm. The numerical solution shows that the temperature and absorbed power density inside the NP are spatially non-uniform due to the decay of the local field away from the metal-dielecric interface. In addition, the absorbed power density along the $z$ axis is different from that along the $x$ and $y$ axes. This happens due to the constructive interference of the electric dipole and quadrupole moments along the axial direction~\cite{bohren_2008_abs_sca_book}. 

However, while the absorbed power density non-uniformity might be significant ($\sim$ 25\% - 50\%, especially, when high-order multipoles are excited), the NP temperature non-uniformity is quite small ($<0.06$\% , see the inset in Fig.~\ref{fig:temp_mie_vs_comsol_40_532}(b)). This is in agreement with simulations performed previously even for strongly asymmetric particles and/or sources~\cite{Baffou-ACSNano-plasmon-heating-shape-2010} which also showed that the reason for the weak non-uniformity is the high thermal conductivity of the metal. This observation allows us to simplify the problem significantly and to obtain an approximate analytic solution of Eq.~(\ref{eq:poissoneq}). Specifically, we replace the spatially non-uniform absorbed power density in Eq.~(\ref{eq:poissoneq}) by its volume average~\cite{Baffou-ACSNano-plasmon-heating-shape-2010}, which can be written as a product of the illumination intensity $I_{\textrm{inc}}$ and the absorption cross-section $C_{\textrm{abs}}$, divided by the volume of the NP, namely, $\bar{p}_{\textrm{abs}}(\omega) \equiv \dfrac{3}{4 \pi a^3} \dfrac{\omega \varepsilon_0}{2}\displaystyle \int d^3 r \varepsilon_m^{\prime\prime}(\omega) |{\bf E}({\bf r},\omega)|^2 = \dfrac{3}{4 \pi a^3} I_{\textrm{inc}} C_{\textrm{abs}}(\omega)$. Then, we integrate Eq.~(\ref{eq:poissoneq}) and apply the Gauss law to convert the volume integral on the left hand side into a surface integral and then integrate over the radius. We obtain~\cite{Baffou-ACSNano-plasmon-heating-shape-2010} \begin{align}\label{eq:solveheatgauss}
\begin{cases}
T(r) = T_{h,0} + \dfrac{\bar{p}_{\textrm{abs}}(\omega)a^2}{3\kappa_h}\left[1 + \dfrac{\kappa_h}{2\kappa_m}\left(1-\dfrac{r^2}{a^2}\right) \right],   & \textrm{ for } r \leqslant a,\\
T(r) = T_{h,0} + \dfrac{ \bar{p}_{\textrm{abs}}(\omega)a^3 }{3 \kappa_h r}, &  \textrm{ for } r \geqslant a.
\end{cases}
\end{align}
where $T_{h,0}$ is the ambient temperature; we hereafter label the ambient properties (zero incident intensity limit) by a subscript 0. Eq.~(\ref{eq:solveheatgauss}) shows that the variation of the temperature inside the NP is at the order of $\sim \kappa_h/\kappa_m$ which is small for common gas and liquid host but can be substantial for dielectric solid host, e.g., for semiconductors~\cite{lide_1995_crc}. As shown in FIG.~\ref{fig:temp_mie_vs_comsol_40_532}(b), the approximate analytic solution Eq.~(\ref{eq:solveheatgauss}) is in excellent agreement with the simulated temperature distribution~\footnote{The simulation domain was equivalent to a cubic box of edge length 4 $\mu$m. Since we set the temperature to 300 K on the outer boundary, the simulated temperature distribution is around 0.4 K (1\%) lower than the approximate analytic solution Eq.~(\ref{eq:solveheatgauss}).}. Thus, it justifies (a-posteriori) the neglect of the spatial dependence of absorbed power density, and more generally, of the results of previous studies of the nonlinear thermo-optic response that relied on the uniform temperature assumption~\cite{Sivan-Chu-Nanophotonics-2017,Gurwich-Sivan-PRE-2017}. 

Hereafter, we denote the particle temperature $T_{\textrm{NP}}$ by its surface temperature, 
\begin{align}\label{eq:T_NP}
T_{\textrm{NP}} = T_{h,0} + \dfrac{\bar{p}_{\textrm{abs}}(\omega)a^2}{3\kappa_h} = T_{h,0} + \dfrac{C_{\textrm{abs}}(\omega)I_{\textrm{inc}}(\omega)}{4 \pi \kappa_h a}.
\end{align}
We also label the solution~(\ref{eq:T_NP}) with $C_{\textrm{abs}}$ obtained from Mie theory and QS approximation by Mie-heat solution and QS-heat solution, respectively. Having established the similarity of the exact numerical solution with the Mie solution, we focus in what follow on the Mie solution and the differences from the QS solution.

\section{Temperature of nanoparticles of different sizes} 
\subsection{A heuristic analysis}
The solution of Eq.~(\ref{eq:T_NP}) implies that, in general, $\Delta T_{\textrm{NP}} \equiv T_{\textrm{NP}} - T_{h,0} \sim C_{\textrm{abs}} / (\kappa_h a)$. For NPs sufficiently small with respect to the wavelength, the electric field within the NP is uniform such that the absorption cross-section scales as $a^3/\lambda$. In this case, the particle temperature scales quadratically with the radius. In contrast, for NPs sufficiently large with respect to the skin depth, the total absorbed power originates mainly from the surface layer of the particle and is approximately proportional to $a^2$. In this case, one expects the particle temperature to be proportional to $a$, i.e., to show a slower (yet monotonic) increase of temperature. However, the numerical results below show that the particle temperature undergoes much more complex variations with the particle size, due to the non-trivial modal response of the particle. We show this through numerical examples.

\subsection{Numeric results - Au NPs}
\begin{figure}[h]
\centering
\includegraphics[width=1\textwidth]{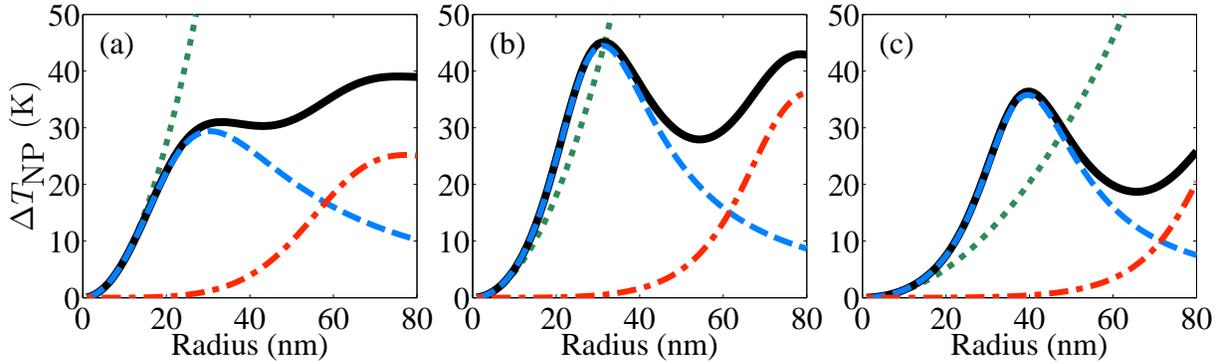}
\caption{\label{fig:temp_vs_size_fixed_I0_oe_532_560_594_nm_lin} (Color online) The temperature rise of gold NP in oil given by the QS-heat method (green dots) and the Mie-heat method (black solid line) as a function of particle radius $a$ under illumination with fixed incoming intensity 40 kW/cm$^2$ at wavelength (a) 532 nm, (b) 560 nm and (c) 594 nm. For the Mie-heat method, the contribution from the electric dipole and electric quadrupole to the particle temperature rise are represented by blue dashed line and red dash-dotted line, respectively. 
}\end{figure}

In FIG.~\ref{fig:temp_vs_size_fixed_I0_oe_532_560_594_nm_lin}, we show the QS-heat and the Mie-heat solutions of Eq.~(\ref{eq:poissoneq}) for gold NPs of different sizes at wavelengths of 532 nm, 560 nm and 594 nm for a fixed incoming intensity of 40 kW/cm$^2$, we also show the contribution from the electric dipole and electric quadrupole modes to the Mie-heat solutions. For sufficiently small particle sizes ($a \lesssim 10$ nm), the Mie-heat solutions show that $\Delta T_{\textrm{NP}}$ increases quadratically with the radius, in agreement with the QS-heat solution; this verifies the validity of the results in Ref.~\cite{Sivan-Chu-Nanophotonics-2017, Gurwich-Sivan-PRE-2017} for such small NPs.

However, for larger NPs, the behaviour is more complicated. Specifically, $\Delta T_{\textrm{NP}}$ increases with the particle radius at a lower and somewhat oscillatory rate for $\lambda = $ 532 nm, see FIG.~\ref{fig:temp_vs_size_fixed_I0_oe_532_560_594_nm_lin}(a). For the longer wavelengths, $\Delta T_{\textrm{NP}}$ exhibits a more interesting behaviour, see FIG.~\ref{fig:temp_vs_size_fixed_I0_oe_532_560_594_nm_lin}(b) and (c). Specifically, for $\lambda = $ 560 nm, the particle temperature rise given by the Mie-heat solution is $\sim$ 10\% {\em higher} than that given by the QS-heat solution for particle radius of 15 nm $\lesssim a \lesssim$ 35 nm. For even larger particles, the size-dependence of $\Delta T_{\textrm{NP}}$ exhibits an oscillatory behaviour which is not captured by the heuristic analysis above. For $\lambda = $ 594 nm, the oscillatory size-dependence of $\Delta T_{\textrm{NP}}$ is similar to the case of $\lambda = $ 560 nm, but the range of sizes for which the Mie-heat temperature is higher than the QS-heat solution is wider and the temperature difference is greater ($\sim 30\%$).

\begin{figure}[h]
\centering
\includegraphics[width=0.6\textwidth]{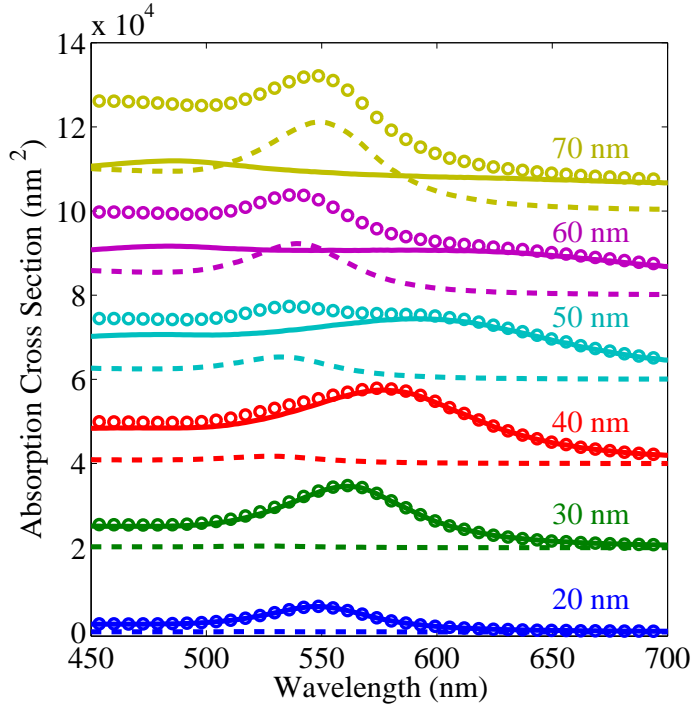}
\caption{\label{fig:abs_size_oe} The (linear) absorption cross-section of gold NPs of difference sizes immersed in oil. The contribution from the electric dipole (solid line) and electric quadrupole mode (dashed line) to the total absorption cross-section (circles) are shown separately.}
\end{figure}

Both the higher particle temperature predicted by the Mie-heat solution and the non-monotonic size-dependence of the particle temperature can be attributed to the contribution of various multipoles to the absorption cross-section. In order to see that, we show in FIG.~\ref{fig:abs_size_oe} and analyze in Appendix~\ref{app:detail_anayls} the contributions to the absorption cross-section from the dipole and quadrupole modes for Au NPs of different sizes. One can see from FIG.~\ref{fig:abs_size_oe} that for particle sizes of $a < 50$ nm, the electric dipole mode dominates the absorption cross-section and that its contribution increases and shifts from $550$ nm to longer wavelengths and then decreases in its quality factor as the particle size increases (see Eqs.~(\ref{eq:dip_res_condi}) and (\ref{eq:dip_res_cabs})). The red-shift of the electric dipole resonance from $550$ nm makes the illumination at longer wavelengths closer to resonance, an effect which is not captured by the QS approximation. This not only explains the faster growing rate of the Mie-heat temperature compared to that of the QS-heat solution for small particle sizes, but also explains that such faster growing rate only occurs for wavelength longer than $550$nm. Secondly, for particle sizes of $a = 50$ nm, the electric quadrupole mode shows up at around $550$nm. As the particle size further increases, the electric quadrupole mode contribution gets stronger, shifts to longer wavelengths and then dominates over the electric dipole mode (see Appendix~\ref{app:detail_anayls}). The further shift and weakening of the electric dipole response, as well as the emergence of the electric quadrupole mode explains the non-monotonic dependence of particle temperature on the particle size for wavelengths longer than $550$ nm.

\subsection{Numeric results - Ag NPs}
\begin{figure}[b]
\centering
\includegraphics[width=0.75\textwidth]{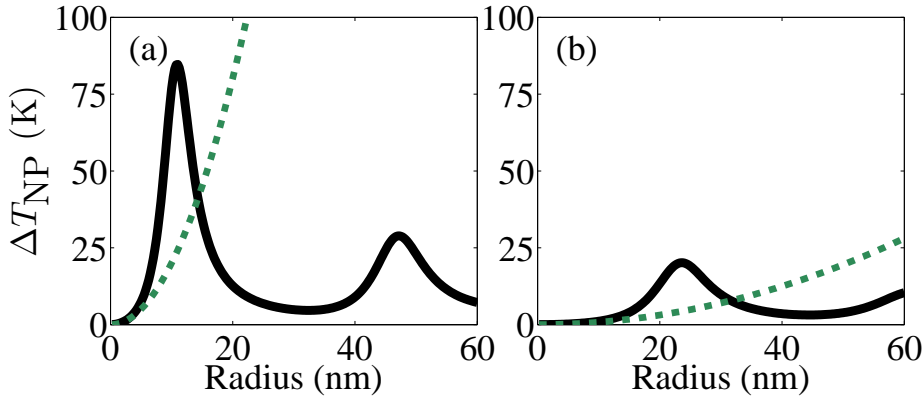}
\caption{\label{fig:temp_vs_size_fixed_I0_ag_420_450_nm_lin} (Color online) The temperature rise of an Ag NP in oil given by the QS-heat method (green dots) and the Mie-heat method (black solid line) as a function of particle radius $a$ under illumination of 40 kW/cm$^2$ at wavelength (a) 420 nm and (b) 450 nm.}
\end{figure}
FIG.~\ref{fig:temp_vs_size_fixed_I0_ag_420_450_nm_lin}(a) and (b) show the comparison of the size-dependent particle temperature between the QS-heat solution and the Mie-heat solution for Ag NP under illumination intensity 20 kW/cm$^2$ at wavelength 420 nm (the plasmon resonance wavelengrth for small Ag spheres in oil) and 450 nm, respectively. The particle size range in which the QS-heat solution agrees with the more accurate Mie-heat solution is now much smaller than $10$ nm. The non-monotonic size-dependence of the Ag NP temperature is again due to the red-shift of the dipole and due to the emergence of the quadrupole resonance, qualitatively similar to the case of Au NPs. The difference in the non-monotonic size-dependence between Ag and Au NPs is caused by the much smaller value of $|\epsilon^{\prime\prime}_m/\epsilon^{\prime}_m|$ for Ag than that of Au.

\section{Discussion}
\begin{figure}[h]
\centering
\includegraphics[width=0.75\textwidth]{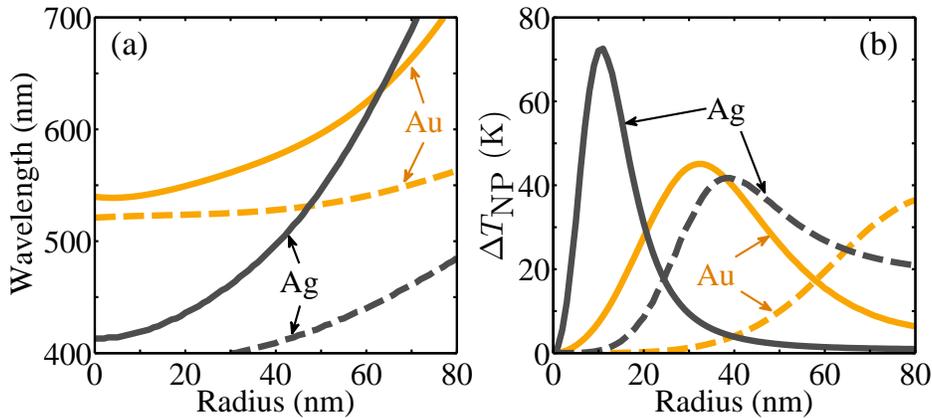}
\caption{\label{fig:dT_wl_ed_eq_au_vs_ag_on_res} (Color online) (a) The resonance wavelength and (b) The particle temperature rise of Au and Ag NPs in oil as a function of particle radius $a$ under illumination of 40 kW/cm$^2$ at the corresponding resonance condition. The solid and dashed lines represents the electric dipole and quadrupole modes, respectively.}
\end{figure}

The oscillatory size-dependence of particle temperature rise shown in the numerical examples above implies that there exists an optimal size of NPs as nano-heating sources for each wavelength. Since the plasmonic resonance shifts to longer wavelengths when the particle size increases, as shown in FIGs.~\ref{fig:abs_size_oe} and~\ref{fig:dT_wl_ed_eq_au_vs_ag_on_res}(a), one can further optimize the performance of the photo-thermal response by choosing the illumination wavelength to match the plasmonic resonance of the metal NP. For example, we show in FIG.~\ref{fig:dT_wl_ed_eq_au_vs_ag_on_res}(b) the size-dependence of the particle temperature rise at the corresponding resonance wavelength. The combination of Figs.~\ref{fig:dT_wl_ed_eq_au_vs_ag_on_res}(a) and~\ref{fig:dT_wl_ed_eq_au_vs_ag_on_res}(b) shows that under the electric dipole (quadrupole) resonance condition the highest temperature rise occurs at $a = 32 \textrm{ nm}\ \&\ \lambda = 564 \textrm{ nm}$ ($a = 86 \textrm{ nm}\ \&\ \lambda = 572 \textrm{ nm}$) for Au and $a = 10 \textrm{ nm}\ \&\ \lambda = 420 \textrm{ nm}$ ($a = 40 \textrm{ nm} \&\ \lambda = 410 \textrm{ nm}$) for Ag NPs. In addition, for the electric dipole (quadrupole) resonance, the temperature rise of Ag NP is higher than that of Au NP only for particle size $a < 20$ nm ($a < 60$ nm). 

The size-dependence of the particle temperature rise under the electric dipole resonance condition (\ref{eq:dip_res_condi}) can be understood by the Pad\'e approximation for the absorption cross-section (\ref{eq:dip_res_cabs}). As the particle size increases, the absorption cross-section (\ref{eq:dip_res_cabs}) first increases with the NP volume ($x^3$ in the numerator), then it is subsequently enhanced by the dynamic depolarization ($x^2$ in the numerator and the denominator) and finally is suppressed due to the radiation damping ($x^3$ in the denominator). Eq.~(\ref{eq:dip_res_cabs}) can also be used to explain the material-dependence of the particle temperature rise. For small particle sizes, the absorption cross-section (\ref{eq:dip_res_cabs}) is {\em inversely} proportional to $\varepsilon_{m}^{\prime\prime}$, so the temperature rise is higher for low loss metal NP, in agreement with previous results~\cite{Baffou-Quidant-Thermo-plasmonics-review-2013,Gurwich-Sivan-PRE-2017}. However, when the particle size increases and becomes large enough so that the radiation damping term ($x^3$ term in the denominator of (\ref{eq:dip_res_cabs})) dominates, the absorption cross-section becomes {\em directly} proportional to $\varepsilon_{m}^{\prime\prime}$, so the temperature rise is higher for lossy metal NPs. The size- and material-dependence of the temperature rise associated with the electric quadrupole resonance are qualitatively similar to those of the electric dipole resonance except that $\Delta T_{\textrm{NP}} \sim a^4$ for small particle sizes (see Eq.~(\ref{eq:cabs_eq_55})).

The oscillatory size-dependence of particle temperature rise shown in FIGs.~\ref{fig:temp_vs_size_fixed_I0_oe_532_560_594_nm_lin} and \ref{fig:temp_vs_size_fixed_I0_ag_420_450_nm_lin} can be used to explain results of various previous experiments. Specifically, the local minimum of the size-dependent fluence threshold for photothermal bubble generation~\cite{Nano_bubble_gen_NP_ACSNano_2010,ps_to_ns_nano_bubble_gen_Langmuir_2014,Baffou_bubble_gen_thr_JPCC.119.28586_2015}.

This study is also an important step towards achieving a quantitative match between experimental results and verifying the thermal effect~\cite{Sivan-Chu-Nanophotonics-2017,Gurwich-Sivan-PRE-2017} being the underlying mechanism responsible for the nonlinear scattering of CW illumination observed in~\cite{Shi-Wei_PhysRevLett.112.017402_2014,Shi-Wei_ACS.Photonics.1.32_2014,Shi-Wei_ACS.Photonics.3.1432_2016}. 

We should mention that our model can be further improved. For example, Meng {\em et al.}~\cite{F.J._plasmonic_oven_ACS_Nano_2017} used the more accurate the two-temperature model~\cite{anisimov_electron_emission_TTM_1974} that accounts for the differences between the electron and lattice temperatures, and accounts for the Kapitza resistance~\cite{Kapitza_Res_RevModPhys.41.48_1969,thm_bdy_res_RevModPhys.61.605_1989,Munjiza_2014,thm_bdy_cond_annurev_matsci_2016,Kapitza-NPs,Kapitza-Au-silicon}. This study also showed that the photo-thermal response can be further improved by carefully adding multiple metal shells. 

Finally, we note that if one is interested in studying the thermal response for temperature rise greater than 100 K, it is necessary to take into account the temperature dependence of the optical and thermal properties of the metal (and its surroundings) when calculating the field and temperature under intense illumination conditions~\cite{Sivan-Chu-Nanophotonics-2017}. A complete temperature-dependent and size-dependent model for the photothermal response will have to be left for a future study.

\section*{Acknowledgement} 
YS and IWU were partially supported by Israel Science Foundation (ISF) grant no. 899/16. IWU was partially supported by Joan and Irwin Jacobs TIX Institute at National Tsing Hua University in Taiwan.

\appendix

\section{Detailed analysis of the size-dependence of the absorption cross-section}\label{app:detail_anayls}
In this Appendix, we analyze in detail the size-dependence of the absorption cross-section of a nanosphere. Based on Mie theory~\cite{bohren_2008_abs_sca_book}, the absorption cross-section can be quantified by the Mie coefficients, denited by $a_n$'s and $b_n$'s. Since we are interested only in metallic NPs, in this analysis we will focus on $a_n$'s, which are responsible for the plasmonic resonance. The electric Mie coefficient $a_n$ is given by~\cite{bohren_2008_abs_sca_book}
\begin{align}\label{eq:an_mie}
a_n &= \dfrac{m\psi_n(mx)\psi^{\prime}_n(x) - \psi_n(x)\psi^{\prime}_n(mx)}{m\psi_n(mx)\xi^{\prime}_n(x) - \xi_n(x)\psi^{\prime}_n(mx)},
\end{align}
where $m = \sqrt{\varepsilon_m/\varepsilon_h}$ is the contrast parameter, $x = k a$ is the size parameter, $k = \sqrt{\epsilon_h}\omega/c_0$ is the wavenumber of the host, while $\psi_n(\rho)$ and $\xi_n(\rho)$ are the Riccati-Bessel functions:
\begin{align}\label{eq:Riccati_Bessel}
\psi_n(\rho) = \rho j_n(\rho),\quad \xi_n(\rho) = \rho h_n^{(1)}(\rho).
\end{align}
By applying the Pad\'e approximants~\cite{Sihvola_pade_to_mie_PhysRevB.94.140301,Sihvola_pade_to_mie_review_2017}, the first electric Mie coefficient $a_1$ is approximated by
\begin{align}\label{eq:pade_a1_33}
a_1^{[3/3]} \approx \dfrac{-i\dfrac{2}{3}\dfrac{\varepsilon_m-\varepsilon_h}{\varepsilon_m+2\epsilon_h}x^3}{1 - \dfrac{3}{5}\dfrac{\varepsilon_m-2\varepsilon_h}{\varepsilon_m + 2\varepsilon_h}x^2-i\dfrac{2}{3}\dfrac{\varepsilon_m - \varepsilon_h}{\varepsilon_m + 2\varepsilon_h}x^3}\equiv-\dfrac{2i}{3}\dfrac{(\varepsilon_m-\varepsilon_h)x^3}{q_1(\varepsilon_m^{\prime},\varepsilon_m^{\prime\prime},\varepsilon_h,x)+iq_2(\varepsilon_m^{\prime},\varepsilon_m^{\prime\prime},\varepsilon_h,x)},
\end{align}
where $r$ and $s$ in $[r/s]$ denote the highest order of the polynomial in the numerator and denominator of the Pad\'e approximated function, respectively. Specifically, 
\begin{align}\label{eq:q1_a33}
q_1(\varepsilon_m^{\prime},\varepsilon_m^{\prime\prime},\varepsilon_h,x) = \varepsilon_m^{\prime}\left(1-\dfrac{3}{5}x^2\right) + 2\varepsilon_h \left(1+\dfrac{3}{5}x^2\right) +\dfrac{2}{3}\varepsilon_m^{\prime\prime}x^3,
\end{align}
and 
\begin{align}\label{eq:q2_a33}
q_2(\varepsilon_m^{\prime},\varepsilon_m^{\prime\prime},\varepsilon_h,x) = \varepsilon_m^{\prime\prime} \left(1 - \dfrac{3}{5}x^2\right) - \dfrac{2}{3}(\varepsilon_m^{\prime} - \varepsilon_h)x^3.
\end{align}

The $x^2$ and $x^3$ terms in the denominator are, respectively, recognized as the dynamic depolarization~\cite{meier_dyn_depolar_1983} and radiation damping~\cite{Wokaun_rad_damp_PhysRevLett.48.957}. The dipole resonance condition can be determined by the zero of the real part of the denominator of $a_1^{[3/3]}$, i.e., $q_1 = 0$, which is satisfied for 
\begin{align}\label{eq:dip_res_condi}
\varepsilon_m^{\prime} = \dfrac{-2\varepsilon_h(1+3x^2/5) - 2\varepsilon_m^{\prime\prime}x^3/3}{1-3x^2/5}.
\end{align}
The quality factor of the resonance is determined by the imaginary part of the denominator of $a_1^{[3/3]}$, i.e., $q_2$. One can clearly see the red-shift and the reduction in the quality factor of the dipole resonance from Eq.~(\ref{eq:dip_res_condi}) and (\ref{eq:q2_a33}) due to the increase of particle size. 

The contribution of the electric dipole to the absorption cross-section can be explicitly written as
\begin{align}\label{eq:cabs_ed_33}
C^{\textrm{ed}[3/3]}_{\textrm{abs}} = \dfrac{6\pi}{k^2}\left[\textrm{Re}(a_1^{[3/3]}) - |a_1^{[3/3]}|^2\right] \approx \dfrac{12\pi}{k^2}\dfrac{\varepsilon^{\prime\prime}_m\varepsilon_h x^3(1+x^2)}{q_1^2+q_2^2}.
\end{align}

The electric dipole absorption cross-section under resonance condition can be obtained by substituting Eq.~(\ref{eq:dip_res_condi}) into Eq.~(\ref{eq:cabs_ed_33}) 
\begin{align}\label{eq:dip_res_cabs}
C^{\textrm{ed}[3/3]}_{\textrm{abs,res}} \approx \dfrac{12\pi}{k^2}\dfrac{\varepsilon_m^{\prime\prime}\varepsilon_h x^3(1+x^2)}{\left[\varepsilon_m^{\prime\prime}\left(1-3x^2/5\right)-2\left(\varepsilon_m^{\prime}-\varepsilon_h\right)x^3/3\right]^2}.
\end{align}

\begin{figure}[h]
\centering
\includegraphics[width=1\textwidth]{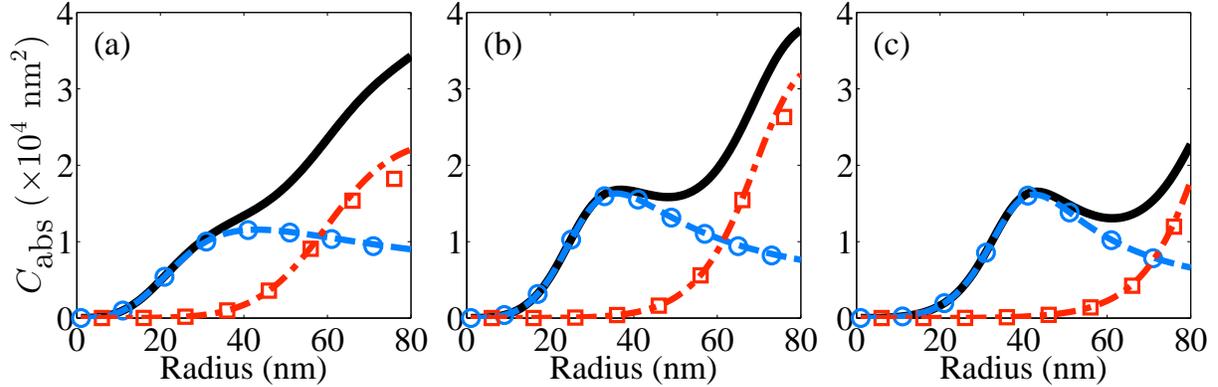}
\caption{\label{fig:cabs_vs_size_fixed_I0_oe_532_560_594_nm_lin_mie_vs_pade} (Color online) The absorption cross-section of gold NP in oil given by the Mie theory (black solid line) as a function of particle radius $a$ at wavelength (a) 532 nm, (b) 560 nm and (c) 594 nm. The contribution from the electric dipole and electric quadrupole to the absorption cross-section (and their Pad\'e approximation) are represented by blue dashed line (blue circles) and red dash-dotted line (red squares), respectively.}
\end{figure}

The Pad\'e approximated second electric Mie coefficient $a_2$ can be obtained similarly~\footnote{Here, higher order terms are required in order to increase the accuracy of the Pad\'e approximate.} 
\begin{align}\label{eq:cabs_eq_55}
a_2^{[5/5]} \approx \dfrac{-\dfrac{i}{15}\dfrac{\varepsilon_m-\varepsilon_h}{2\varepsilon_m+3\varepsilon_h}x^5}{1+\dfrac{5}{7}\dfrac{\varepsilon_h}{2\varepsilon_m+3\varepsilon_h}x^2-\dfrac{5}{1323}\dfrac{\varepsilon_m^2+30\varepsilon_m\varepsilon_h-45\varepsilon_h^2}{\varepsilon_h(2\varepsilon_m+3\varepsilon_h)}x^4-\dfrac{i}{15}\dfrac{\varepsilon_h}{2\varepsilon_m+3\varepsilon_h}x^5}.
\end{align}
After some lengthy algebra, one can see from $C_{\textrm{abs}}^{\textrm{eq}[5/5]} = \dfrac{10\pi}{k^2}\left[\textrm{Re}(a_2^{[5/5]}) - |a_2^{[5/5]}|^2\right]$ that $C_{\textrm{abs}}^{\textrm{eq}[5/5]}$ changes with the particle size in a similar way to $C_{\textrm{abs}}^{ed[3/3]}$ except that $C_{\textrm{abs}}^{\textrm{eq}[5/5]} \sim x^5$ for small particle size.

In FIG.~\ref{fig:cabs_vs_size_fixed_I0_oe_532_560_594_nm_lin_mie_vs_pade}, we show a comparison of size-dependent $C_{\textrm{abs}}$ of a single Au NP between the Mie theory and the Pad\'e approximation. The Pad\'e approximation shows excellent agreement with the Mie theory for particle size $a \lesssim 60$ nm.

\end{document}